\documentclass[12pt]{article}
\usepackage{amsmath}
\usepackage{amssymb}
\usepackage{times}
\usepackage{color}
\usepackage{epsfig,graphics}
\begin{document}
\pagestyle{plain}
\hsize = 6.5 in
\vsize = 8.5 in
\hoffset = -0.75 in
\voffset = -0.5 in
\baselineskip = 0.29 in

\def \rvX{{\bf X}}
\def \rvY{{\bf Y}}
\def \rvz{{\bf Z}}
\def \rvT{{\bf T}}
\def \rvD{\hbox{\boldmath $\Delta$}}
\def \rt{\hbox{\boldmath $\xi$}}
\def \rphi{\hbox{\boldmath $\phi$}}
\def\vx{{\bf x}}
\def\vn{{\bf n}}

\title{Mesoscopic Biochemical Basis of Isogenetic
Inheritance and Canalization: Stochasticity, 
Nonlinearity, and Emergent Landscape}
\author{Hong Qian\\[8pt]
Department of Applied Mathematics, 
University of Washington\\
Seattle, WA 98195, USA.\\[8pt] 
Hao Ge\\[8pt]
Beijing International Center for 
Mathematical Research\\ 
and Biodynamic Optical Imaging Center\\ 
Peking University, Beijing, 100871, PRC.} 

\maketitle

\abstract{Biochemical reaction systems in mesoscopic
volume, under sustained environmental chemical gradient(s), can have
multiple stochastic attractors.  Two distinct mechanisms are known
for their origins: ($a$) Stochastic single-molecule events, such as
gene expression, with slow gene on-off dynamics; and ($b$) nonlinear
networks with feedbacks. These two mechanisms yield different 
volume dependence for the sojourn time of an attractor. As in the
classic Arrhenius theory for temperature dependent transition
rates, a landscape perspective provides a natural framework 
for the system's behavior.  However, due to the nonequilibrium 
nature of the open chemical systems, the landscape, and the attractors it represents, are all themselves {\em emergent 
properties} of complex, mesoscopic dynamics. In terms of the 
landscape, we show a generalization of Kramers' approach 
is possible to provide a rate theory.  The emergence of 
attractors is a form of self-organization in the mesoscopic 
system; stochastic attractors in biochemical systems such
as gene regulation and cellular signaling are naturally 
inheritable via cell division. Delbr\"{u}ck-Gillespie's 
mesoscopic reaction system theory, therefore, provides a 
biochemical basis for spontaneous isogenetic switching and canalization.
}

\section{Introduction}

Epigenetic inheritance at the cellular level preserves certain
phenotypes through cell divisions [\cite{leibler_05}]. Since the
definition of ``epigenetic'' means the inheritability is not due to
genes, i.e., DNA sequences, the epigenetic phenomenon must be a
biochemical process [\cite{ptashne_2007,spudich_76}]. DNA and histone
modifications, gene regulations by transcriptional factors,
signaling networks, and metabolic pathways are all parts of cellular
biochemistry.  Current research chiefly focuses on the 
``code'' of epigenetic inheritance in terms of DNA methylation 
and/or histone acetylation [\cite{turner_2000,jones_2001,zhang_cocb_07}].

\begin{figure}[t]
\[
\psfig{figure=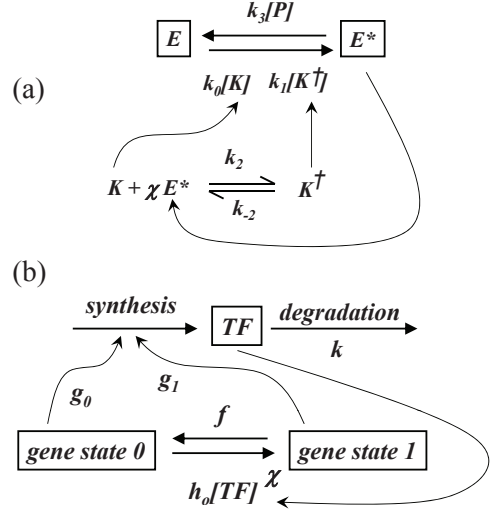,width=2.75in}
\]
\caption{A biochemical signaling network of
phosphorylation-dephosphorylation cycle (PdPC)
and gene expression regulatory network for a 
self-regulating gene are essentially isomorphic.  
(a) An enzyme $E$ can be phosphorylated to become 
$E^*$, catalyzed by a kinase $K$. 
Dephosphorylation of $E^*$ is catalyzed by
phosphatase $P$. The activity of $K$, however, is  
modified through binding $\chi$ copies of
$E^*$: $K^{\dag}$ denotes $K(E^*)_{\chi}$ complex.
(b) The expression of a transcription factor (TF)
is regulated by the very TF, in monomer or dimer form
($\chi=1,2$).  The TF is degradated with rate $k$. $g_1$ and
$g_0$ are the respective biosynthesis rates when the TF 
is on and off the gene.  If $g_1<g_0$, the TF is a 
repressor; and if $g_1>g_0$ the TF is an activator.
[Redrawn based on Fig. 14 of \cite{gqq}.]}
\label{fig1}
\end{figure}

    While the gene expressions and their regulations are
a central component of epigenetic processes, it is less certain what
the roles of cellular signaling networks, or metabolic
networks, are.  Even more importantly, are different
gene expressions the {\em cause} of the epigenetic phenomenon,
or consequences of the dynamics of a larger intracellular 
biochemical network as a whole? In the present
paper, we advance a theory for biochemical reaction systems 
in a mesoscopic volume [\cite{qian_jsp}]. Taking a broader 
perspective that is rooted in stochastic, nonlinear 
dynamical systems [\cite{qian_iop}], we illustrate that it is
likely that the code for epigenetic inheritance is {\em
distributive} [\cite{hopfield_84}]. Using two classes
of biochemical networks (see Fig. \ref{fig1}) as illustrations,
we investigate mesoscopic, nonlinear biochemical reaction 
networks with multiple {\em stochastic attractors} [\cite{wolynes_pnas_05}].  More importantly, we shall 
show that two very different types of mesoscopic 
bistabilities exist: the {\em stochastic bistability}
which has no macroscopic counterpart
[\cite{elston_bj_01,wolynes_pnas_02,wolynes_pre_05,arkin_pnas_05,
wolynes_nobel_09,qian_bj_10}], 
and the {\em nonlinear bistability}
which exhibits deterministic bistability when the system's 
volume tends to macroscopic scale [\cite{qian_jrsi_09,qian_prl_09,liang_qian,shi_qian_11}].

    There is a growing awareness of mesoscopic biochemical
bistability [\cite{wang_xj_05,dodd_cell_07,Turcotte_pnas,
Arup_cell,cooper_pnas_09}].  Multistability in a
mesoscopic chemical reaction system can be mathematically 
represented by the Delbr\"{u}ck-Gillespie processes (DGP) 
whose probability distribution function follows chemical 
master equation [\cite{delbruck}] and whose exact
stochastic trajectories can be obtained by the widely
known Gillespie algorithm [\cite{arkin_98,gillespie}]. 
This approach has found numerous applications in recent studies.
For an introduction to this chemical reaction system
theory, see [\cite{qian_iop,qian_jsp,liang_qian,
arkin,wilkinson,qian_book}].  A review by \cite{qian_bishop}
is particularly accessible.

    In principle, a biochemical reaction system in a small volume
of the size of a cell, on a very large time scale (years, hundreds
years), will have a steady state probability distribution which
reflects the continuous jumping among the multiple stochastic
attractors: regions with high local probability [\cite{ge_qian_jrsi,qian_prl_09,ao_2004}].
In terms of this {\em stationary distribution}, the dynamics of a
mesoscopic biochemical system can be cogently visualized and
even quantified by a landscape representation [\cite{wolynes,ao_2004,wang_pnas_08,hugang,ge_qian}].

    It will be illustrated in this paper that multistability is not
only stable against noise, i.e., robust, but also could be 
readily inherited during the process of cell volume change and division. Within a cell
cycle, it is not necessary to directly control the concentration
of a biochemical species: As soon as cellular concentrations are
deviated from their locally most stable values, the dynamics will
cause them to spontaneously relax back as long as the deviations are
within the limit of the basin of the attraction (stable state).
Indeed, this restoring phenomenon should be regarded as a form of
``self-organization''.  Such state naturally has stability
and robustness.  On the larger landscape scale, the system is
``digital''. This kind of ``inheritable code'' is not only
distributive but also dynamic [\cite{hopfield_84}], in contrast to
Watson-Crick basepairng.

    Some mathematical analysis of the DGP is at
 the foundation of our current landscape theory [\cite{qian_jsp,qian_iop}].  In Sec. \ref{on_lands} we offer some
discussions. Intuitive and appealing as it is, the justification
of using the stationary probability distribution as the landscape
involves subtle mathematical ideas which deserves
further investigations [\cite{ge_qian_jrsi,ge_qian}].

\section{Biosynthesis of Self-regulating Repressor with 
Slow On-and-off Gene Fluctuations}

    We first consider a simple model for the biosynthesis
and degradation of a repressor with stochastic gene expression,
which is regulated by the repressor.  The canonical kinetic scheme
for this model, neglecting the intermediate stage of mRNA, is in
Fig. \ref{fig1}b.  For simplicity, we assume that the repressor-gene
binding rate $h(n)=h_on$ which involves a monomer, and $g_0>g_1$.
See \cite{elston_bj_01}, \cite{wolynes_pnas_02} \cite{wolynes_pre_05}, and \cite{wolynes_pnas_05} for extensive studies of this model, and related systems with
self-activation dimer ($h(n)\propto n(n-1)$, $g_1>g_0$). 
We choose this model to demonstrate bistability due to
slow, nonadiabatic fluctuations in the gene state.  This is a
stochastic effect due to single-molecule behavior
[\cite{bai_pnas_99,wolynes_nobel_09}] which disappears in the
macroscopic Law of Mass Action [\cite{wolynes_pre_05,qian_pccp_09}]
(see Methods).

\subsection{Stochastic bimodality and bistability}

    It is generally accepted that, in biochemistry,
{\em noise-induced bistability} means a small kinetic system has
bimodal distribution while its macroscopic counterpart has
only uni-stability [\cite{wolynes_pre_05,arkin_pnas_05,qian_bj_10}].
Usually, the meaning of ``macroscopic counterpart'' is defined
as the same chemical or biochemical reaction systems at
same concentrations.  For a large volume, the concentrations
are deterministic variables; but in a mesoscopic volume, the
copy numbers fluctuate. This ``correspondence principle'' is
consistent with the experimental practices: In the past most
biochemical experiments on gene expression were carried out with
DNA measured in concentrations [\cite{PvH_07}].

\begin{figure}[t]
\[
\includegraphics[width=2.2in,angle=270]{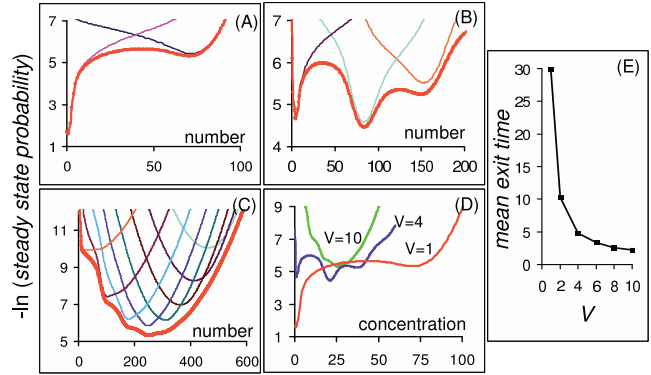}
\]
\caption{Steady state probability distributions for the repressor
molecule numbers in mesoscopic systems with different volume $V$ but
same DNA concentration $x_t=1$. (A) One gene with $V=1$. The red
curve is the repressor distribution.  The two thin curves are
$p^{ss}(n|m)$ with $m=1,0$. Model parameters are taken from
\cite{wolynes_pre_05}. (B) Four genes with $V=4$. The thin curves
are for $p^{ss}(n|m)$ with $m=4,3,2$. (C) Ten genes with $V=10$. 
(D) Distributions for the concentration of the repressor protein 
as functions of increasing $V$ approaching to the macroscopic
prediction with steady state at $y^*=27.8$ (see Methods). (E)
The mean exit time for the all-off, low-transcription state,
defined when possible as the well at $n=Mg_1/k=(g_1x_t/k)V$,
decreases with $V$. }
\label{v_dep_4_sb}
\end{figure}

    With increasing volume of the a reaction system ($V$) 
and the copy numbers DNA ($M$), stability of one of the
two states decreases while the other increases.  For a
sufficiently large $V$, the bistability disappears all
together [\cite{wolynes_pre_05,qian_pccp_09}].
We call such bistability {\em stochastic bistability}.
Fig. \ref{v_dep_4_sb} shows how the stationary probability
distributions change with the increasing $V$ and DNA copy number
$M$ while keeping its concentration $x_t=M/V=1$.

In agreement with the visual landscapes in Figs.
\ref{v_dep_4_sb}A-D, \ref{v_dep_4_sb}E
also shows how the mean sojourn time for the all-off, 
low transcription
state ($m=0, n\le 80$ in Eq. \ref{the_cme_4_srg}) decreases
as a function of $V$.\footnote{The
landscape representation in the nonadiabatic analysis is
valid quantitatively:  The transition
rates between any two states $a$ and $b$, $k_{ab}$ and $k_{ba}$,
satisfy $k_{ab}/k_{ba}=p_b^{ss}/p_a^{ss}$ where $p^{ss}_x$ is
the probability of state $x$.  Let's assume the shape
of each peak is approximately Gaussian. Then the logarithms of
the peak values are $\ln(p_a^{ss}/\sqrt{2\pi}\sigma_a)$ and
$\ln(p_b^{ss}/\sqrt{2\pi}\sigma_b)$,
where $\sigma$'s are the variances of the Gaussian distributions.
Thus the ``energy difference'' between the two wells
$E_b-E_a=-\ln(k_{ab}/\sigma_b)+\ln(k_{ba}/\sigma_a)$, or
$\ln(k_{ab}/k_{ba})=-(E_b-\ln\sigma_b)+(E_a-\ln\sigma_a)$.  The
right-hand-side are the ``free energy'' difference of the two
wells.}  This is in sharp contrast to the nonlinear bistability
(Fig. \ref{fig:switchtime}) we shall discuss next.

\section{Phosphorylation Dephosphorylation Cycle with
Nonlinear Feedbacks}

    A gene regulatory network with dimeric activator also exhibits
bistability [\cite{elston_bj_01,wolynes_pnas_05,shi_qian_11}], 
but by a different mechanism.  To illustrate this, 
and also to broaden the scope of our discussions, we shall 
consider a cellular phosphorylation dephosphorylation 
cycle (PdPC) signaling network in Fig. \ref{fig1}a.  In
particular, we shall consider the case of positive feedback 
with a dimer ($\chi=2$). The same analysis can be applied 
to the gene expression system with $\chi=2$ and $g_1>g_0$
in Fig. \ref{fig1}b. It is easy to show that for 
$\chi=0,1$, the cellular signaling network in 
Fig. \ref{fig1}a, with slow fluctuating kinase
activity, exhibits stochastic bistability
[\cite{qian_bj_10,qian_pccp_09}]. The theory we 
present here is general to both types of 
biochemical networks in terms of nonlinear chemical
kinetics.

The class of networks in Fig. \ref{fig1}a has been widely
implicated, such as in Src family kinase membrane signaling [\cite{cooper_qian}], Rab 5 GTPase in endocytic pathway [\cite{liguangpu}], {\it Xenopus oocytes} regulation
for cell fate [\cite{ferrell_sci}], and long-term neural 
memory [\cite{wang_xj_05,cooper_pnas_09}].
The network has been studied in
\cite{ferrell_chaos,qian_prl_05,qian_prl_09,qian_bj_10}.
The detailed kinetic scheme is given in Eq. \ref{fig3a} 
for which a DGP is uniquely specified (see Methods):  
\begin{eqnarray}
      && E+K^{\dag}+ATP 
   \overset{k_1}{\underset{k_{-1}}{\rightleftharpoons}}
       E^*+K^{\dag}+ADP, 
\label{fig3a}\\
      && K+2E^* 
   \overset{k_2}{\underset{k_{-2}}{\rightleftharpoons}}
      K^{\dag},  \ \ \
      E^* + P 
      \overset{k_3}{\underset{k_{-3}}{\rightleftharpoons}}
       E+P+Pi.
\nonumber
\end{eqnarray}

\begin{figure}[t]
\[
\psfig{figure=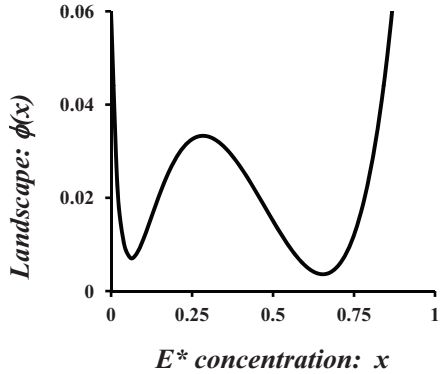,width=2.5in,angle=270}
\]
\caption{The landscape for the PdPC kinetic system 
in Eq. \ref{fig3a}, $\phi(x)$, according to 
Eq. \ref{phi_x}, using parameters $x_t=1$,
$\alpha=43$, $\beta=10$, $\epsilon=0.01$ and $\delta=0.5$.  
From Eq. \ref{fixed_pt}, the steady states are  
$x_1^*=0.05$ (stable), $x_2^*=0.632$ (stable), 
and $x_3^*=0.368$ (unstable).} 
\label{fig2}
\end{figure}

	Note that the stochastic Delbr\"{u}ck-Gillespie approach 
is not an alternative to the traditional enzyme kinetic modeling 
(Eqs. \ref{the_ode}, \ref{fixed_pt}).  When copy numbers 
in a chemical reaction system are large, a Delbr\"{u}ck-Gillespie
process (DGP) automatically yields the deterministic dynamics predicted by the traditional Law of Mass Action [\cite{qian_book,qian_bishop}].  One of the most important
predictions of the DGP theory is a dynamic landscape
for the system, as shown in Fig. \ref{fig2} as well as 
Figs. \ref{v_dep_4_sb}A-D.  Such a landscacpe can only be 
rigorously defined in a stochastic model; it can be
computed using a chemical master equation.

\section{Time Scales, Emergent Landscape and Implications to
Epigenetics}
\label{on_lands}

\subsection{Three time scales of cellular dynamics}

	Double-well landscapes shown in Figs. \ref{v_dep_4_sb} and \ref{fig2} suggest multiple time scales in the biochemical
dynamics.  In fact, there are three distinct time scales 
in such systems.  Note that while fluctuations modify the
``down-hill'' deterministic kinetics, they lead to 
``up-hill'' dynamics which is impossible in a macroscopic
system.  The time for ``barrier crossing'', however, is 
extremely slow in comparison to the down-hill kinetics.

Therefore, the fast time scale in the system is the 
individual biomolecular reactions in Fig. \ref{fig1}.  
For the present work, they are given in terms of the rate
parameters $k$'s in Fig. \ref{fig1}a and $f,g,h,k$ in Fig. \ref{fig1}b (or equivalently, the $\alpha,\beta,\epsilon,\delta$ 
in Eq. \ref{pdpcwfb_2}.)
Millisecond are not unreasonable, even though individual
biochemical reactions inside a cell could be much faster
or slower.

The middle time scale is the relaxation kinetics of a network 
to its steady states, as illustrated in the 
Fig. \ref{fig2} by the downhill dynamics.  Note that the
very existence of a steady state (an attractor), or 
steady states, is a consequence of ``self-organization'' 
of a complex reaction network.

The slow time scale is the transition rates between the 
two basins of attraction, or ``wells''.  Both the middle
(deterministic) and slow (stochastic) time scales are 
{\em emergent properties} of the biochemical network. 
Following \cite{gqq} we shall denote them molecular 
signaling time scale (MSts), biochemical network time 
scale (BNts), and cellular evolution time scale (CEts), 
respectively.

\begin{figure}[h]
\[
\psfig{figure=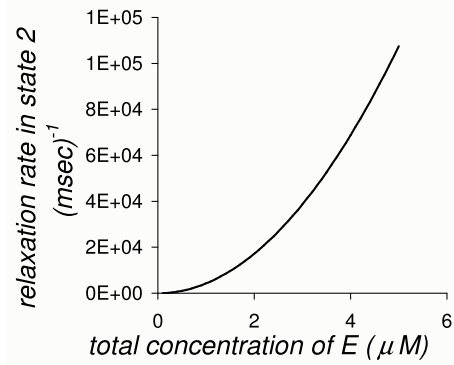,width=2.25in,angle=270}
\]
\caption{Relaxation kinetics in the state 2, the well on
the right in Fig. \ref{fig2} corresponding to state $x_2^*$,
as a function of the total concentration of $E$, $x_t$, according to Eq. \ref{rs}.  The parameter used are
$\alpha=4.3 \mu sec^{-1}\mu M^{-2}$,
$\beta=10 msec^{-1}$, $\epsilon =1 msec^{-1}\mu M^{-2}$
and $\delta=0.5 msec^{-1}$.
}
\label{fig:r2}
\end{figure}

   Again taking the signaling system in 
Fig. \ref{fig1}a (also Eq. \ref{fig3a}) as an example.
For biochemically
realistic situations, $\epsilon\ll\alpha$.  We thus have the landscape
given in Eq. \ref{phi_x} simplified into
\[
    \phi(x) = x_t\ln (x_t-x)-x\ln\left[\frac{(\alpha x^2+\delta)(x_t-x)}
        {\beta x}\right]
\]
\begin{equation}
-2\sqrt{\frac{\delta}{\alpha}}\arctan
    \left(\sqrt{\frac{\alpha}{\delta}}x\right)
        +2x.
\label{phix}
\end{equation}
The parameters $\alpha,\beta,\delta$ in Eq. \ref{phix} 
define the MSts.  We can now use the model to investigate the 
role of $\alpha,\beta$ and $\delta$ on the BNts and CEts.
The BNts is given by the $r_1$ and $r_2$ in Eq. \ref{rs}, and 
the CEts is given by the $T_{1\rightarrow 2}$ and 
$T_{2\rightarrow 1}$ in Eqs. \ref{T12_exact}-\ref{T21}.

    The BNts changes with total concentrations of the regulators
$E$ inside the system, as well as the concentrations of other factors.
Fig. \ref{fig:r2} shows how in the simple model the time-scale
decrease, i.e., rate increases, with the total concentration of $E$,
$x_t$.  Within less than one order of magnitude change of $x_t$,
from 1 to 6$\mu M$, the relaxation rate in the state 2, i.e., the
well on the right in Fig. \ref{fig2}, increase by a factor
of 100.  Eq. \ref{rs} confirms that there is a square dependence of
the relaxation rate to the concentration.

\begin{figure}[t]
\[
\psfig{figure=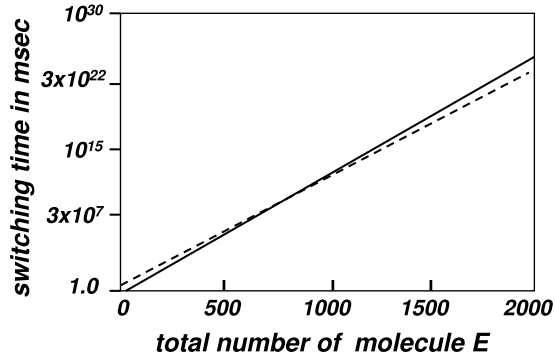,width=2.in,angle=270}
\]
\caption{The switching times between two stable states,
$T_{1\rightarrow 2}$ (dashed line) and $T_{2\rightarrow 1}$ (solid
line) increase with the volume $V$ in nonlinear bistability
(according to Eqs. \ref{T12_exact}). $x_t=0.1 \mu M$, $\alpha=4.3
\mu sec^{-1}$, $\beta=10 msec^{-1}$, $\epsilon =1 msec^{-1}$ and
$\delta=0.5 msec^{-1}$.  A cell has a volume at about 10-20
femtoliter.  With concentration $x_t=0.1 \mu M$, there are 600-1200
number of total molecules.  This gives the switching time around
$3\times 10^4$ to $10^{12}$ sec.  That is 10 hours to thirty
thousand years!  Therefore, with rather rapid individual biochemical
reaction rates $\alpha$, $\beta$, $\epsilon$ and $\delta$, the
emergent epigenetic states can be extremely stable.}
\label{fig:switchtime}
\end{figure}

    The CEts is extremely sensitive to the number of
molecules in the biochemical system. Fig. \ref{fig:switchtime} 
shows that with the MSts on the order of milli-
and microsecond, and with concentration of $E$ on the order of
micromolar , the transition times between the two states in Fig.
\ref{fig2} can be as long as thirty thousand years!  Hence, the
stability of the emergent attractors could be extremely 
robust against spontaneous concentration fluctuations (i.e., intrinsic noise) in the system.  We also notice that 
both transition rates decrease with the $V$.  This will be 
explained in Sec. \ref{sec:on_land} below.

More interesting biologically, we note 
that the range of 700-1000 copy number of $E$ 
corresponds to a time range of 10 hours to 30 years. 
In yeast, \cite{oshea} have reported that 
the copy numbers for most of the transcription factors 
are centered arround $2^{10}=1024$ per cell.

\subsection{Epigenetic inheritance and canalization on the CEts}

\begin{figure}[t]
\[
\psfig{figure=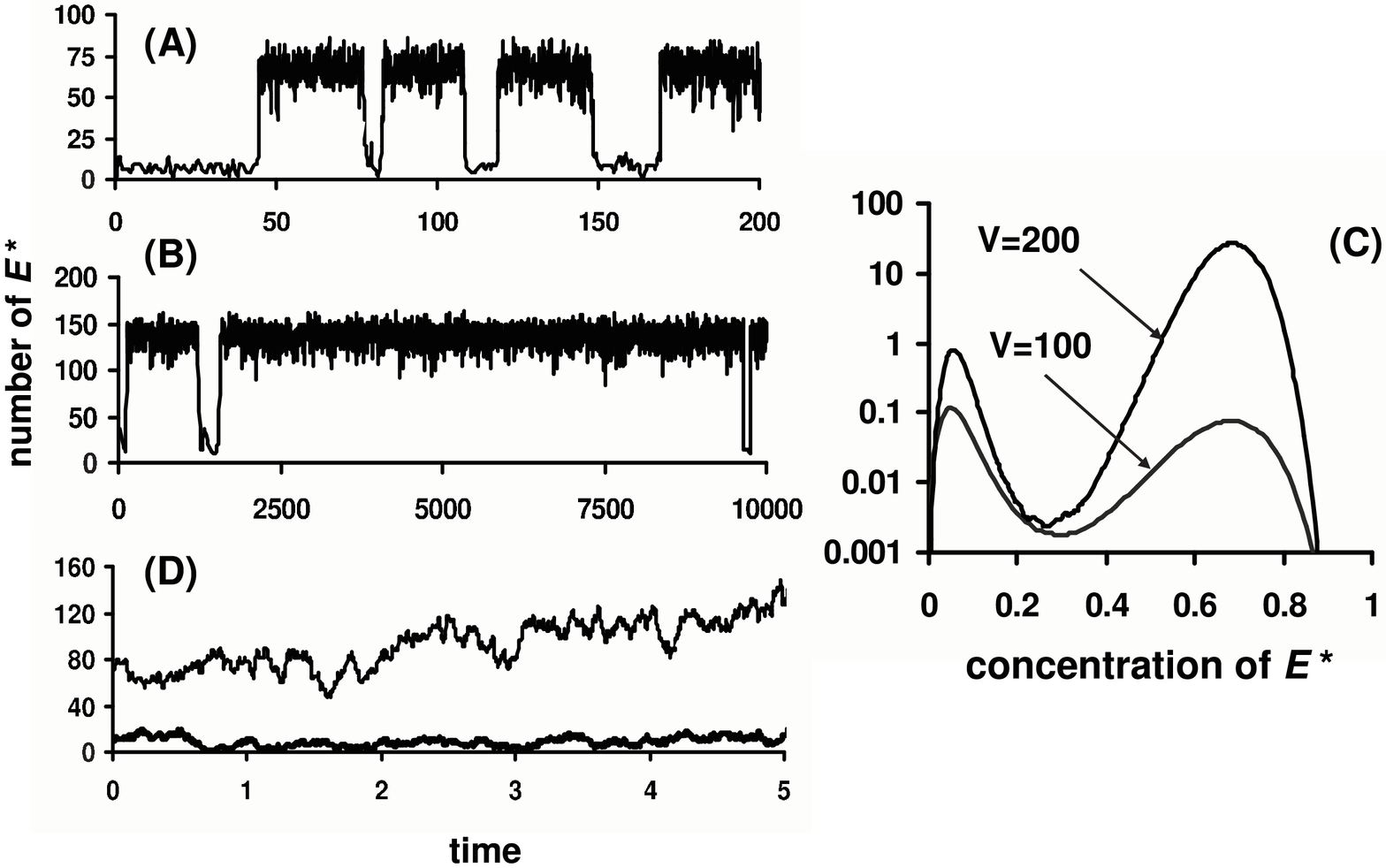,width=2.1in,angle=270}
\]
\caption{Nonequilibrium steady state fluctuation of the
number of phosphorylated $E^*$ as functions of time, with $V=100$
in (A) and $V=200$ in (B).  The steady state distributions of the
number of $E^*$ for the $V=100,n_t=100$ and $V=200,n_t=200$ are shown in (C).
The parameters used are $\alpha=43,\beta =10,\epsilon=0.01$ and
$\delta=0.5$.  (D) The volume $V$ and $n_t$ are increased by a factor of
2 from $V=100, n_t=100$ to $V=200,n_t=200$ within the time 0 to 5.  One
sees that the two attractors are well separated in the volume doubling
process.}
\label{fig3}
\end{figure}

    Let us consider two replica of a mesoscopic biochemical
reaction system in a laboratory, for example one of those in
Fig. \ref{fig1}a which do not involve
gene expression.  If the two systems have same total $E$ but different
initial values for $E^*$, one near zero and one
near the total $E$, then these two systems settle to the two different
attractors.  In the time scale much shorter than the evolutionary
transitions, the numbers of $E^*$ fluctuate around the $n_1^*$ and
$n_2^*$, respectively, or equivalently around the $c_1^*$ and $c_2^*$ if
the volume of systems do not change.  (If the volumes are changing,
then the fluctuations are around the $c_1^*$ and $c_2^*$, but not
$n_1^*$ and $n_2^*$!) However, at the time scale greater than the cellular
evolution, there will be transitions between the two attractors.
This is shown in Fig. \ref{fig3}A.  The probability distribution for
the concentration of $E^*$ is shown in Fig. \ref{fig3}C.

    Fig. \ref{fig3}B shows the identical reaction system, except 
its volume and total $E$ are twice as large (keeping the concentration invariant).
What we observe from the Fig. \ref{fig3}C
is that the ``concentration'' distribution for $E^*$ in the two cases
have essentially the same locations for the peaks and trough.   This implies
that if the size of the biochemical reaction system increases in the time scale
sufficiently short, then the identities of the attractors can be preserved.

    The stable state of the system is not only stable against intrinsic
noise, but also could be readily transferred to the two daughter
cells. During the cell cycle, the concentrations of biochemical
substances might become approximately one half of the
original value, in the extreme cases due to cell volume increase,
and the system deviates from its corresponding stable state.
The kinetic law of these two daughter systems is just the
same as their mother cell except they have not relaxed to any of the
stable states. As we have shown previously that the relaxation scale
is not very large, so they will spontaneously relax back to the same
corresponding stable state as long as not leaving an basin.
Many previous work of epigenetics always searched for a stable
chemical substance like DNA, which could be self-propagated and
inherited to the daughter cells, while here we give another
alternative possibility that the code of epigenetic is at the
dynamic level of the whole biochemical network.

    Further, there are clear upper and lower bounds for the rate of
volume increase:  it can not be too large such that the instantaneous
changing concentration is outside the basin of an attractor; it can
not be too slow such that it is on the cellular evolution time
scale. Surely, the stability of epigenetic code is weaker than that
for a stable chemical substance and it is more flexible facing the
influence of the fluctuating environment, but it is suffient for a
normal cell, a chemical system, to survive and inherit even in a fluctuating environment.

    Fig. \ref{fig3}D shows precisely two of such simulations.  Consider
the volume, and the total $E$, double within the time of 5 units.
This is a duration much shorter than the cellular evolution time.
The top and bottom traces are the number of $E^*$ from two
simulations.  At the end of the doubling, if each system is divided
into two, both ``daughter systems'' will also inherit the biochemical
state of the ``mother system''.  The biochemistry of a 
mesoscopic reaction system is inheritable!   The epigenetic 
stability, i.e., canalization [\cite{zhang_cocb_07}] could
be directly related to the CEts.

\subsection{Mesoscopic stochastic dynamics on a landscape}
\label{sec:on_land}

    The foregoing discussion clearly illustrates
the power of the ``landscape'' perspective in visualizing and
representing the global dynamics in bistable systems. We
see that
for stochastic bistability, at least one of the ``barrier heights''
decrease with volume $V$ (Fig. \ref{v_dep_4_sb}), while for
nonlinear bistability, the both barrier heights increase with the
volume (Figs. \ref{fig:switchtime} and \ref{fig3}).

    What determines the entire landscape?  Since it is defined on
the space for all possible concentrations and/or copy numbers of all
the molecular species in the reaction system, itself can not be
determined by the concentrations and copy numbers.  Rather, it is
determined by the all possible molecules involved and their
interaction/reaction rate constants.  In other words, biochemical
reaction networks.  Since the molecular interaction/reaction rate
constants are properties of molecular structures, which in turn is
determined by the primary sequences in the case of proteins, we
conclude that the landscape, conceptually, is encoded in the DNA
sequence, together with the extracellular environment including the
volume $V$, but is independent of the expression patterns of
transcription factors.  They are the consequences of a 
biochemical reaction system's dynamics 
[\cite{wolynes_pnas_02}].

    There are many similarities between the energy 
landscape for a protein in equilibrium [\cite{wolynes}] 
and the landscape for an open mesoscopic chemical 
system in a nonequilibrium steady state 
[\cite{ao_2004,qian_arpc_07,gqq}].
We, however, want to emphasize a key difference:
Recall that an energy landscape exists {\em a priori} for
a dynamical protein \cite{wolynes}.
The open-chemical systems are fundamentally different 
in this respect [\cite{qian_jpc_06,qian_arpc_07,gqq}].  
Specifically, for any finite volume
$V$, a mesoscopic reaction system has a 
stationary probability distribution for
the number of copies of all its biochemical species, $p_V(\vn)$,
where $\vn=(n_X,n_Y,n_Z,\cdots)$ are the copy numbers of the
molecules $X,Y,Z$ etc.  It can be shown that such a distribution
can be expressed as
\begin{equation}
    p_V(\vn) = \exp\left[ -V\phi(\vx)+\phi_1(\vx) 
       + \frac{\phi_2(\vx)}{V} + \cdots\right], 
\label{land_scape}
\end{equation}
where $\vx=\vn/V$.  Furthermore, 
it can be shown that the function $\phi(\vx)$
is a meaningful landscape for the complex dynamics of the nonlinear biochemical
system.  That is, the dynamics always goes ``down the hill'' of $\phi(\vx)$ [\cite{hugang,ge_qian}], though usually not 
by the steepest descent path.
As we have seen, the landscape provides an very useful organizational
device for thinking about cellular biochemical dynamics at widely
different time scales, ranging from individual signaling reactions to
cellular phenotype switching.

Eq. \ref{land_scape} shows that the stationary distribution
$p_V(\vx)$ changes with $V$.  When the $V$ becomes macroscopic
volume, the probabilities are concentrated only at the global
minima of the
function $\phi(\vx)$.   However, with increasing $V$, $-(1/V) \ln
p_V(\vn)$ approach to the function $\phi(\vx)$ which is defined on
the entire space of $\vx$.  It is an important insight that such a
landscape exist and it is independent of the system's volume, as
long as the volume is reasonably large [\cite{hugang,ge_qian}].

    The nonlinear bistability, therefore, is the mesoscopic
manifestation of a double well in the $\phi(\vx)$.  Its macroscopic
counterpart has two stable steady states.  However, stochastic
bistability is something very different: The bimodal distribution
only exists when the $V$ is sufficiently small; when $\phi_1$ and
$\phi_2$ in Eq. \ref{land_scape} contribute to the $p_V(\vn)$. 
$\phi(\vx)$ has only a single
well.  With increasing volume: the barrier in the
nonlinear bistability increases, while that of stochastic
bistability decreases.

    The emergent landscape of cellular interaction network
dynamics and the landscape for protein dynamics are
fundamentally different.  The lack
of detailed balance due to the open chemical nature of the former
gives rise to the cycle flux underneath the landscape
[\cite{wang_pnas_08,gqq}].  The cycle flux makes the 
landscape non-local.
When such a flux is sufficiently strong (i.e.,
mathematically characterized by the emerging of complex eigenvalue
and eigenvectors), a synchronized dynamics arises
[\cite{qian_qian_prl,ge_mbs_08}]. The emergence of synchronized
dynamics requires an entirely new kind of phenomenology.  In the
macroscopic classical
world, this is the birth of oscillatory behavior and wave phenomena
that have ruled classical engineering for a century.

\section{Discussion}

\subsection{Adaptive landscape}

    While the concept of energy landscape becoming a very useful
term in protein dynamics, the concept of adaptive landscape in
evolutionary dynamics is still highly controversial [\cite{hartl_pnas_10}].
We believe one of the main reasons for this situation is that the landscape
in the latter, as the landscape in the present work, is an emergent entity,
which is not given {\it a priori}.   There are two issues related to
this important distinction: (1) The mathematical existence of such a
landscape in a general, nonlinear stochastic dynamics 
which does not have
detailed balance; and (2) How is such a landscape, even exists,
related to the dynamics, both deterministic and stochastic.  The
most nontrivial issue here is the logical relationship between the
landscape and the dynamics: It is rather clear that in systems 
with detailed balance, the landscape exists {\it a priori} and the dynamics is a {\bf\em consequence} of the landscape.  However, 
for system without detailed balance, the dynamics, as 
defined by the reaction networks and all the individual rate constants, define the overall dynamics {\bf\em as well as} 
define the landscape.  Dynamics and the landscape have
{\bf\em correlations} but no {\bf\em causality}.  Hence, 
logically, it is
not correct to view the dynamics as a consequence of the landscape.
Nevertheless, the landscape is still a very useful device
to understand and characterize the overall dynamics. The concept of
emergent landscape, thus, should only be understood in this
``historically retrospective'' sense.

\subsection{Genocentric epigenetic inheritable memory and
a possible alternative}

Currently, methylation of DNA is considered to be the leading
candidate for epigenetic inheritance.  Even though this
mechanism is not based on Watson-Crick basepairing, its function is
still intimately dependent on the discovery of 1950s. Namely, the
``code'' is still a part of an extended, modified DNA structure.
While certainly methylation is a part of the ``whole picture'' of
epigenetic regulation, such a view might be too genocentric and too
limited.  It is not unlikely that the code of epigenetic
inheritance is dynamics rather than static, distributive rather than
localized: It is an emergent property of the whole system rather
than depend on a very few number of substances or regulatory
mechanisms. The emergent landscape of a mesoscopic chemical reaction
system share certain features with the content addressable memory
proposed by J.J. Hopfield years ago, but eliminated the technical
need for detailed balance [\cite{hopfield_84}].

We would like to point out the fundamental difference of our
proposal is a non-genome, pure biochemical based epigenetic
mechanism.  While DNA methylation is a part of this picture, 
our proposed
mechanism moves the focus away from DNA basepair recognition and
memory, and shift it to biochemical networks.   The memory in our
model, in principle, can be independent of DNA.
We understand that immunological diversity and memory
has now been shown to be DNA based [\cite{tonegawa_83}].  
Still, the digital ``quantal'' nature of immunity at
individual cell level has been noted [\cite{smith_06}].
Also, in the field of neuroplasticity, cellular 
mechanism for memory has been proposed to be very similar
to our PdPC model [\cite{wang_xj_05,cooper_pnas_09}].
A biochemical based memory is too natural to be completely
neglected by evolution.   

At single cell level, bimodality has been regularly
observed in diverse cell types. Recently, for example, 
\cite{xie_08} demonstrated in their Fig. 1C bimodal 
expression level of lactose permease in {\em E. coli.} 
corresponding to the two states of a bacteria cell, 
uninduced or induced by lactose analog 
TMG (methyl-b-D-thiogalactoside). \cite{zhang} in 
their Figure 4 reported bimodal Raman spectra as 
an indicator for DNA fragmentation in apoptosis 
of DAOY cell line (human brain tumor medulloblatoma),
and in Figures 6 and 7, \cite{xu} showed bimodal 
FITC-Annexin V protein in apoptosis of U2OS 
cell line (human osteosarcoma).

\subsection{What is a pathway, and what are cross-talks?}

    The concept of biochemical regulatory pathway
is widely accepted in the molecular cellular biology.  In
general applications, a pathway provides a sequential events of
activation/deactivation in terms of the
regulatory/signaling proteins.  However, from our biochemical
reaction system perspective, the state of a cell, as a
stochastic attractor, is defined by the
states of {\bf\em all} the regulatory/signaling proteins
[\cite{ge_mbs_08}].  Hence, from a more rigorous theoretical
standpoint, one needs to know, when a protein $Y$ is activated, what
are the states of its upper-stream and down-stream proteins, $X$ and
$Z$. In this sense, the ``pathway view'' of cellular signaling is
similar to the ``structural pathway'' view of protein conformational
change.  While it is useful, it can be mis-leading [\cite{cui_06}].

    Almost universally in the discussion of cellular
signaling, the concept of ``certain pathway leading to certain
response'' is an established language.  But knowing a great deal of
``cross-talks'' in signaling pathways [\cite{zhang_ncb_08,rama_06}],
this linear thinking is incorrect.  In fact, a response is really
a changing state of a cell, be it proliferation versus growth arrest,
apoptosis versus senescence.  So they should not be associated with
{\bf\em only} one pathway or another. Rather, they are different
states of {\bf\em entire integrated network}.  It is true, by
``activating certain pathway'' while keeping other the same, one
might promote a particular state, but that does not imply the
response is only associated with that pathway.

Now if we take the view these responses are different 
``states'' of a cell, then many pathways are involved.  
And the most important things about biological complexity 
is how many these stable states
are ``available'' in the system [\cite{wolynes_pnas_02}]. 
More states also means more possible transitions between them (whether a transition actually occurs is an issue of time scales), 
and thus the complexity is associated with
multi-stability. This view is consistent with the definition of
cellular complexity by possible number of responses to combinatorial
stimuli [\cite{rama_06}].  If there is only one state, then the system always goes back to somewhere it starts, then no 
complexity.

The key idea here is not to associate a particular cellular 
response to a particular pathway; but rather activating 
particular pathway promotes certain response which is 
defined as state change.

\subsection{Living matter: Mesoscopic open chemical system
as a biochemical machine}

    \cite{Laughlin} have
discussed the possible new phenomena at mesoscopic
scale which they called the middle way. Others has
asked ``what is and how is living matter different from
the three (or five) known states of matters from physics
[\cite{cele_07}].  The study of mesosopic chemical systems
offers some insights:

1)  If a chemical system is too large (for example, grinding all the
cells into a single tube without the small volume of a cell), then
the chemistry is different.  It completely loses the possibility
of multi-stability at that CEts [\cite{qian_prl_09}], which is
a defining feature of complex dynamics [\cite{qian_pccp_09}].
The size of biological cells might indeed be a consequnce of
living matters possessing mesoscopic complexity.

2) This idea is generally understood but its implication has not
been widely appreciated: A living matter has to be in continuous
exchange of materials, with chemical gradient, with its environment;
The driven nature of a living matter is completely different from
the classical way of thinking a ``matter'' which is being in
isolation. To define a living matter, one can not completely
separate the system from its environment:
This is the origin of systems view of holistic biology.

3) The classical thesis of ``irreversibility'' of Boltzmann is to
understand the spontaneous processes leading to equilibrium. This is
a different problem as that suggested by Schr\"{o}dinger in his
``what is life''.   The former is to understand the macroscopic
irreversibility in a system with Newtonian mechanics, while the
latter is a completely different problem; Life is an open system: To
a first-order approximation, it is an isothermal system sustained by
an active input and output of chemicals with Gibbs free energy
difference.  The irreversibility of such a system is self-evident.
Hence, essential thesis in nonequilibrium physics of living matter
is not that of Boltzmann, but more quantitative understanding of how
such chemically driven systems give rise to complex ``living''
behavior such as self-organization and inheritability.

4) The significance of the bistability arising from the
Delbr\"{u}ck-Gillespie is not that it has two different 
``phenotypical'' states for relatively low and relatively 
high levels of ``inducer'', borrowing the language from 
Lac operon [\cite{xie_08}], but that both states co-exist 
for an range of intermediate level of inducer!  This is
reflected by the distribution in this intermediate
range being bimodal.  Such a realization was emphasized 
in protein folding by \cite{lattman_pnas_93}.
This realization can have important implications:
A pre-cancerous state might already exist ``on the other side
of the mountain'', which is encoded in our genome [\cite{aoping_med_hy}].

\section{Methods}

\subsection{Self-regulating gene and stochastic
bistability}

    We consider the coupled birth-death process
for the gene regulatory network in Fig. \ref{fig1}b with $\chi=1$.
We note that in a macroscopic biochemical experiment the rate of
protein synthesis per DNA is $g_i$ $(i=0,1)$.  Furthermore, we
denote the concentration of the DNA with repressor $x=m/V$ and the
concentration of the repressor $y=n/V$.  Then
\begin{eqnarray}
   &&  \frac{dp(m,n)}{dt}
\nonumber\\ 
   &=& g(m)p(m,n-1)\ +\ (n+1)kp(m,n+1) 
\nonumber\\
   &+& \frac{h_on}{V}(M-m-1)p(m-1,n)
\label{the_cme_4_srg}\\
   &-& \left(g(m)+nk+\frac{h_on}{V}(M-m)+mf\right) p(m,n)
\nonumber\\
   &+& (m+1)fp(m+1,n),
\nonumber
\end{eqnarray}
where $g(m)=g_0(M-m)+g_1m$,
$M=x_tV$ is the copy number of DNA.  This model,
in the limit of $V\rightarrow\infty$, yields the
Mass Action kinetic equation
\begin{equation}
    \frac{dx}{dt} = h_oy(x_t-x)-fx, \  \ \
    \frac{dy}{dt} = g_0(x_t-x) + g_1x
            - ky.
\label{ode_4_srg}
\end{equation}
Eq. \ref{ode_4_srg} has two roots but one
in the interval $(0,x_t)$.  For parameters
$g_0=80,g_1=1,k=1,h_o=0.007,f=0.1,k=1.0,x_t=1$, we
have the steady state $x^*=0.66$ and $y^*=27.8$.

    Using nonadiabatic approximation, we have
$p(n|m)$ being a Poisson distribution with mean $g(m)/k$.
Then the 2-d model is reduced to 1-d with birth rates
$h_o(M-m)g(m)/(kV)$ and death rate $mf$.  The
stationary distribution for the 1-d model can be
analytically studied following \cite{qian_bj_10}.

    Mathematically, one can understand the problem as an
eigenvalue perturbation:  The Markov operator involved has the block
structure [\cite{qian_pccp_09}]:
\begin{equation}
    \left(\begin{array}{cc}
        \mathbf{L_0}-\mathbf{h}  &  \mathbf{f} \\
                  \mathbf{h} & \mathbf{L_1}-\mathbf{f}
            \end{array}\right),
\end{equation}
in which $\mathbf{h}$ and $\mathbf{f}$ are small and treated as a
perturbation. The unperturbed operator has a degenerated eigenvalue
zero, with eigenvectors on the left $(\mathbf{1},\mathbf{1})$ and
$(\mathbf{1},-\mathbf{1})$, and on the right
$(\mathbf{p}_0,\mathbf{p}_1)^T$ and
$(\mathbf{p}_0,-\mathbf{p}_1)^T$. $\mathbf{p}_0$ and $\mathbf{p}_1$
are Poisson distributions with mean $\overline{n}_0 =g_0/k$ and
$\overline{n}_1=g_1/k$, which are the stationary distributions for
the nonperturbed problem. Now for the perturbed system, it is easy
to verify that zero is still an eigenvalue $\lambda_0=0$; however,
there is also a nonzero, smallest eigenvalue
\begin{equation}
    \lambda_1 =  \frac{\left(\begin{array}{cc}
           \mathbf{1},-\mathbf{1}
        \end{array}\right)}{2}
        \left(\begin{array}{cc}
        -\mathbf{h}  &  \mathbf{f} \\
                  \mathbf{h} & -\mathbf{f}
            \end{array}\right)
        \left(\begin{array}{c}
           \mathbf{p}_0 \\ -\mathbf{p}_1
        \end{array}\right) = -(h_o\overline{n}_0+f).
\end{equation}
Furthermore, the approximated right eigenvectors associated with
$\lambda_0$ and $\lambda_1$ are precisely
\begin{equation}
    \left(\frac{f}{h_o\overline{n}_0+f}
    \mathbf{p}_0,\frac{h_o\overline{n}_0}{h_o\overline{n}_0+f}
    \mathbf{p}_1\right)
    \ \ \textrm{ and }  \ \
    (\mathbf{p}_0,-\mathbf{p}_1).
\end{equation}
A more accurate approximations for the two eigenvectors can be
obtained, if needed, by carrying through the first-order
perturbation calculations [\cite{shi_qian_11}]. 
The eigenvectors for the $\lambda_1$ has
a {\em nodal decomposition} which yields two connected domains with
positive and negative values.\footnote{The mathematical theory for
the bistability proceeds with the following argument: If one treats
the $\lambda_1$ as a small parameter, then for system with
$\lambda_1=0$, the stochastic dynamics can be reduced to two
independent subsystems, each with a unique stationary distribution.
Therefore, with the error on the order of $\lambda_1$, both
eigenvectors for the $\lambda_0=0$ and $\lambda_1\neq 0$ are
linear combinations of the two stationary distributions defined on the
subspaces. This explains the origin of the two-state
behavior on the slow time scale (CEts) \cite{qian_jrsi_09}.}
This provides a rigorous mathematical definition for the
two states of the system.  Since all the other eigenvalues
$\lambda_i\gg\lambda_1$ $(i\ge 2)$, the dynamics within each of the
two domains are on a different time scale, and fast equilibrated.
The remaining slow dynamics corresponds precisely to a two state
system [\cite{qian_jrsi_09}] with transition rates $h$ and $f$. The
nonadiabaticity plays a decisive role in this problem.\footnote{The
problem appears very similar to the quantum mechanical eigenvalue
perturbation with degeneracy. However, it is worth pointing out that
in quantum mechanics one computes the eigenvalues which serves as
the ``energy landscape'' in Heitler-London theory; here we compute
the eigenvectors as the ``energy landscape''.  This distinction has
been noted by \cite{qian_pccp_09}. }

\subsection{Chemical master equation and landscape $\phi(x)$}

The PdPC with feedback in Eq. \ref{fig3a}
is the same as that in Fig. \ref{fig1}a with $k_0=0$.  With rapid
binding $K+2E^*\rightleftharpoons K^{\dagger}$ and assuming $\frac{k_{-2}}{k_2}
\gg [E^*]^2$, we have the following autocatalytic, nonlinear chemical
reaction system:
\begin{equation}
    E+2E^* \overset{\alpha}{\underset{\epsilon}{\rightleftharpoons}} 3E^*,
\ \
    E^* \overset{\beta}{\underset{\delta}{\rightleftharpoons}} E,
\label{pdpcwfb_2}
\end{equation}
in which $\alpha = \frac{k_1k_2}{k_{-2}}[K][ATP]$, $\beta =k_3[P]$,
$\epsilon=\frac{k_{-1}k_2}{k_{-2}}[K][ADP]$, and $\delta=k_{-3}[P][Pi]$,
where $[K]$ and $[P]$ are the concentrations of the kinase and the
phosphatase. One can see that this kinetic model is intimately related to
the Schl\"{o}gl model which has been the prototype of nonlinear
chemical bistability [\cite{qian_book,qian_jrsi_09}].

    Let $x$ be the concentration of $E^*$, then the macroscopic
kinetic equation for (\ref{pdpcwfb_2}) in terms of the Law of Mass
Action is
\begin{equation}
    \frac{dx}{dt} =\alpha x^2(x_t-x)-\beta x
                -\epsilon x^3 +\delta (x_t-x),
\label{the_ode}
\end{equation}
where $x_t$ is the total concentration of the $E$ and $E^*$,
assumed to be constant in the system.  The dynamic
has three positive fixed points.  For most biochemical applications
$\epsilon\ll\alpha$ and $\delta\ll\beta$.  Hence, one can
approximately have the three steady states
\[
    x_1^* = \frac{\delta x_t}{\beta},\ \
    x_2^* = \frac{x_t+\sqrt{x_t^2-4\beta/\alpha}}{2},
\]
and
\begin{equation}
      x_3^* = \frac{x_t-\sqrt{x_t^2-4\beta/\alpha}}{2}.
\label{fixed_pt}
\end{equation}
Hence, when $\alpha x_t^2 >4\beta$, there is bistability.
$x_1^*$ and $x_2^*$ are stable steady states, and $x_3^*$ is unstable.  The basins of attraction for $x_1^*$ and 
$x_2^*$ are $[0,x_3^*)$ and $(x_3,\infty)$,
respectively.  The corresponding linear relaxation 
rate for the steady state $x_i^*$, $(i=1,2)$, is
$r_i = 3\alpha\left(x^*_i\right)^2-2\alpha x_t x_i^*+\beta$.  That is,
\begin{equation}
    r_1 =\beta \ \textrm{ and } \
    r_2 =\alpha x_2^*(2x_2^*-x_t).
\label{rs}
\end{equation}

    The chemical master equation for the system in 
Eq. \ref{pdpcwfb_2} is
\begin{eqnarray}
    && \frac{dp(n)}{dt} =
\nonumber\\
    && \left(\frac{\alpha}{V^2}(n-1)(n-2)+\delta\right)(n_t-n+1)p(n-1)
\nonumber\\
    &+& \left(\beta+\frac{\epsilon}{V^2}n(n-1) \right)(n+1)p(n+1)
\nonumber\\
    &-& \left[\left(\frac{\alpha}{V^2}n(n-1)+\delta \right)(n_t-n)
               \right.
\nonumber\\
    && \left. +\left(\beta+\frac{\epsilon}{V^2}(n-1)(n-2) \right)n\right]p(n),
\end{eqnarray}
where $n_t$ is the total number of $E$ and $E^*$ molecules, and $V$ is the
volume of the mesoscopic system.
The steady state probability distribution for the
number of $E^*$ is
\begin{equation}
    p^{ss}(n) = C\prod_{k=2}^n
        \frac{[\alpha k(k-1)+\delta V^2](n_t-k)}
        {\left[\beta V^2+\epsilon k(k-1)\right](k+1)},
\end{equation}
where $C$ is a normalization factor.  If $V$ and $n_t$ are large,
but their ratio is hold constant, then one can develop an approximated
formula for the probability density function for the concentration
$x=n/V$:
\begin{equation}
        f^{ss}(x) \propto e^{-V\phi(x)},
\end{equation}
where the $\phi(x)$
\begin{equation}
    \phi(x) = x_t\ln (x_t-x)-x\ln\left[\frac{(\alpha x^2+\delta)(x_t-x)}
        {(\beta+\epsilon x^2) x}\right]
\label{phi_x}
\end{equation}
\[
    - 2\sqrt{\frac{\delta}{\alpha}}\arctan
    \left(\sqrt{\frac{\alpha}{\delta}}x\right)
        +2\sqrt{\frac{\beta}{\epsilon}} \arctan
        \left(\sqrt{\frac{\epsilon}{\beta}}x\right).
\]
One can check the extrema of $\phi(x)$ by setting its derivative being zero:
\begin{equation}
    \frac{d\phi(x)}{dx} = -\ln\frac{(\alpha x^2+\delta)(x_t-x)}
            {(\beta+\epsilon x^2)x} = 0.
\end{equation}
We see that the extrema of $\phi(x)$ are precisely the roots of
Eq. \ref{the_ode}, $x_1^*,x_2^*$ and $x_3^*$: The ratio
in the logarithm being 1 corresponds to the right-hand-side of
Eq. \ref{the_ode} being 0.

\subsection{The switching time}

    The mean time for switching from $x_1^*$ attractor to $x_2^*$
attractor can be analytically computed for 1-d model.
For discrete case, this
formulae has been widely used in the various lattice hopping models
with birth-death processes \cite{vanKampen}:
\[
    T_{1\rightarrow 2} = \sum_{n=0}^{n_1^*}p^{ss}(n)
        \sum_{m=n_1^*+1}^{n_2}\frac{1}{w_mp^{ss}(m)}
\]
\begin{equation}
        +\sum_{n=n_1^*+1}^{n_2^*-1}p^{ss}(n)
        \sum_{m=n+1}^{n_2^*}\frac{1}{w_mp^{ss}(m)}.
\label{T12_exact}
\end{equation}
where $n_i^*=[x_i^*V]$, and
\[ 
  u_n = \left(\frac{\alpha}{V^2}(n-1)(n-2)+\delta\right)(n_t-n+1), \]
\[
    w_n = \left(\beta+\frac{\epsilon}{V^2}(n-1)(n-2)\right)n.
\]
When $V\rightarrow\infty$, $w_m\sim w\left(\frac{m}{V}\right)V$
where $w(x)=\beta x+\epsilon x^3$, and
$p^{ss}(n)\sim (C/V)e^{-V\phi\left(\frac{n}{V}\right)}$,
we have (see Appendix)
\begin{equation}
    T_{1\rightarrow 2} \approx \frac{2\pi e^{V(\phi(x_3^*)-\phi(x_1^*))}}
        {w(x_3^*)\sqrt{-\phi^{''}(x_1^*)\phi^{''}(x_3^*)}}.
\label{T12}
\end{equation}
and similarly,
\begin{equation}
    T_{2\rightarrow 1}\approx \frac{2\pi e^{V(\phi(x_3^*)-\phi(x_2^*))}}
        {w(x_3^*)\sqrt{-\phi^{''}(x_2^*)\phi^{''}(x_3^*)}}.
\label{T21}
\end{equation}
This result, as expected, is very similar to Kramers formula for
overdamped barrier crossing in the energy landscape $\phi(x)$ with
``frictional coefficient'' being $1/w(x_3^*)$.   Note that at
$x_3^*$, $w(x_3^*)=u(x_3^*)$; hence a symmetric expression for the
``fractional coefficient'' is $2/(w(x_3^*)+u(x_3^*))$.

\section{Acknowledgements}

We thank Peter Wolynes and X. Sunney Xie for many
helpful discussions. HQ was partially supported by 
NSF grant EF0827592 (PI: Dr. H. Sauro).


\appendix
\section{Mean Switching Time for 1-D CME}

The mean time for switching from $x_1^*$ attractor to $x_2^*$
attractor can be analytically computed.  For discrete case,
according to \cite{vanKampen}, let
\[  
 u_n = \left(\frac{\alpha}{V^2}(n-1)(n-2)+\delta\right)(n_t-n+1), 
\]
\[
  w_n = \left(\beta+\frac{\epsilon}{V^2}(n-1)(n-2)\right)n,
\]
we have
\[
        T_{1\rightarrow 2} =
    \sum_{n=0}^{x_1^*V}p^{ss}(n)
     \sum_{m=x_1^*V+1}^{x_2^*V}\frac{1}{w_mp^{ss}(m)}
\]
\begin{equation}
        +\sum_{n=x_1^*V+1}^{x_2^*V-1}p^{ss}(n)
       \sum_{m=n+1}^{x_2^*V}\frac{1}{w_mp^{ss}(m)}.
\end{equation}
$w_m\approx w\left(\frac{m}{V}\right)V$,
where $w(x)=\beta x+\epsilon x^3$, and $p^{ss}(n)\approx
(C/V)e^{-V\phi\left(\frac{n}{V}\right)}$, we have
\begin{eqnarray}
    T_{1\rightarrow 2} &\approx&
  \sum_{n=0}^{x_1^*V}\frac{e^{-V\phi(\frac{n}{V})}}{V}
  \sum_{m=x_1^*V+1}^{x_2^*V}\frac{1}{w(\frac{m}{V})
    e^{-V\phi(\frac{m}{V})}}
\nonumber\\
   && +\sum_{n=x_1^*V+1}^{x_2^*V}\frac{e^{-V\phi(\frac{n}{V})}}{V}\sum_{m=n+1}^{x_2^*V}\frac{1}{w(\frac{m}{V})e^{-V\phi(\frac{m}{V})}}
\nonumber\\
   &\approx&\sum_{n=0}^{x_1^*V}e^{-V\phi(\frac{n}{V})}\int_{x_1^*}^{x_2^*}\frac{1}{w(x)e^{-V\phi(x)}}dx
\nonumber\\
  && +\sum_{n=x_1^*V+1}^{x_2^*V-1}e^{-V\phi(\frac{n}{V})}\int_{\frac{n}{V}}^{x_2^*}\frac{1}{w(x)e^{-V\phi(x)}}dx
\nonumber\\
  &\approx&V\int_{0}^{x_1^*}e^{-V\phi(x)}dx\int_{x_1^*}^{x_2^*}\frac{1}{w(x)e^{-V\phi(x)}}dx
\nonumber\\
   &&+V\int_{x_1^*}^{x_2^*}e^{-V\phi(y)}\int_{y}^{x_2^*}\frac{1}{w(x)e^{-V\phi(x)}}dxdy.
\nonumber
\end{eqnarray}
Applying Laplace's method, one has
\[
    \int_{0}^{x_1^*}e^{-V\phi(x)}dx\approx\frac{\sqrt{2\pi}
    e^{-V\phi(x_1^*)}}{2\sqrt{V\phi^{''}(x_1^*)}},
\]
\[  \int_{x_1^*}^{x_2^*}\frac{1}{w(x)e^{-V\phi(x)}}dx
    \approx\frac{{\frac{\sqrt{2\pi}}{w(x_3^*)e^{-V\phi(x_3^*)}}}}
        {\sqrt{-V\phi^{''}(x_3^*)}},
\]
\[  \int_{y}^{x_2^*}\frac{1}{w(x)e^{-V\phi(x)}}dx
    \approx
    \left\{\begin{array}{ll}
       \frac{{\frac{\sqrt{2\pi}}{w(x_3^*)e^{-V\phi(x_3^*)}}}}{\sqrt{-V\phi^{''}(x_3^*)}}  & 
        x_1^*\leq y\leq x_3^*,
\\[8pt]
     -\frac{\frac{1}{w(y)e^{-V\phi(y)}}}{V\phi^{'}(y)} & 
         x_3^*\leq y\leq x_2^*.
\end{array}\right.
\]
Then 
\begin{equation}
    T_{1\rightarrow 2} \approx
    \frac{\pi e^{V(\phi(x_3^*)-\phi(x_1^*))}}{w(x_3^*)
    \sqrt{-\phi^{''}(x_1^*)\phi^{''}(x_3^*)}}
\end{equation}
\[
   +V\int_{x_1^*}^{x_3^*}e^{-V\phi(y)}dy\frac{{\frac{\sqrt{2\pi}}{w(x_3^*)e^{-V\phi(x_3^*)}}}}{\sqrt{-V\phi^{''}(x_3^*)}}
\]
\[
  +V\int_{x_3^*}^{x_2^*}e^{-V\phi(y)}-\frac{\frac{1}{w(y)e^{-V\phi(y)}}}{V\phi^{'}(y)}dy.
\]
Furthermore, $$\int_{x_1^*}^{x_3^*}e^{-V\phi(y)}dy\approx
\frac{\sqrt{2\pi}e^{-V\phi(x_1^*)}}{2\sqrt{V\phi^{''}(x_1^*)}},$$
hence
\begin{equation}
T_{1\rightarrow 2}\approx \frac{2\pi
e^{V(\phi(x_3^*)-\phi(x_1^*))}}{w(x_3^*)\sqrt{-\phi^{''}(x_1^*)\phi^{''}(x_3^*)}}
-\int_{x_3^*}^{x_2^*}\frac{1}{w(y)\phi^{'}(y)}dy,\nonumber
\end{equation}
in which the second term is comparable to
$\int_{x_3^*}^{x_2^*}\frac{1}{u(y)-w(y)}dy$ which is the time
relaxing from the unstable fixed point $x_3^*$ to $x_2^*$. It should
be neglectable when the noise is present.

Finally, when $V\rightarrow\infty$, we have the switching time in
terms of $\phi(x)$:
\begin{equation}
    T_{1\rightarrow 2}\approx \frac{2\pi
e^{V(\phi(x_3^*)-\phi(x_1^*))}}{w(x_3^*)\sqrt{-\phi^{''}(x_1^*)\phi^{''}(x_3^*)}},
\label{time_1_to_2}
\end{equation}
and
\begin{equation}
    T_{2\rightarrow 1}\approx \frac{2\pi
e^{V(\phi(x_3^*)-\phi(x_2^*))}}{w(x_3^*)\sqrt{-\phi^{''}(x_2^*)\phi^{''}(x_3^*)}}.
\end{equation}

\subsection*{Comparison with Kramers' theory} 

The Kramers theory
considers a diffusing particle that obeys the Fokker-Planck
equation
\[
    \frac{\partial p(x,t)}{\partial t}
    = \frac{1}{\eta} \left(
        \frac{\partial}{\partial x} [U^{'}(x)p(x,t)]
        +k_BT\frac{\partial^2}{\partial x^2} p(x,t)
        \right).
\]
For energy function $U(x)$ with two energy wells, say well $a$ and
well $b$, he derived \cite{vanKampen} the rate constant for
transition from $a$ to $c$ by crossing barrier $c$, known as the
celebrated Kramers' formula:
\begin{equation}
    k_{a\rightarrow b} = \frac{\sqrt{\mu_a\mu_c}}{2\pi\eta}
    e^{(U_a-U_c)/k_BT},
\end{equation}
where $\mu_a$ and $\mu_c$ are the curvatures of the
energy function $U(x)$ at $a$, the well, and $c$, the
transition state.  Similarly one has $k_{b\rightarrow a}$.
Then their ratio, which relates the populations in the
two wells at equilibrium:
\[
    \frac{k_{a\rightarrow b}}{k_{b\rightarrow a}}
    =
    \sqrt{\frac{\mu_a}{\mu_b}}
    e^{-\frac{U_b-U_a}{k_BT}} = e^{-\frac{F_b-F_a}{k_BT}},
\]
where $F_a = U_a+k_BT\ln\sqrt{\mu_a}$ is the free energy of
energy well $a$.

\subsection*{Comparison with diffusion approximation to CME} 

The result in Eq. \ref{time_1_to_2}, as far as we know, 
is new.  In the past,
analysis of barrier crossing in the CME, in the limit of large $V$,
have been based on diffusion approximation to the CME.  In
that approach, one derives a one-dimensional Fokker-Planck equation
for large $V$:
\[
\frac{\partial p(x,t)}{\partial t}
    =-\frac{\partial}{\partial x}
    (u(x)-w(x))p(x,t)
\]
\begin{equation}
    +\frac{\partial^2}{\partial x^2}
    \left(\frac{u(x)+w(x)}{2V}\right)p(x,t).
\end{equation}
Here, it suggests a potential
\begin{equation}
    \psi(x)=-2\int^x \frac{u(x)-w(x)}{u(x)+w(x)}\ dx,
\end{equation}
which is different from our
\begin{equation}
    \phi(x)=-\int^x \log\frac{u(x)}{w(x)}\ dx.
\end{equation}
$\phi(x)$ is the correct one while $\psi(x)$ could give incorrect 
result for the $T_{1\rightarrow 2}$ and $T_{2\rightarrow 1}$. 
See [\cite{qian_jrsi_09}] for an extensive discussion.

\section{Transition Time in Nonadiabatic Gene Switching}

We consider the coupled birth-death process with probability distribution
$\left[p_0(n),p_1(n)\right]$, where $0$ and $1$ represent the
gene states, with and without bound transcription factor, and $n$
represent the copy number of the transcription factor which
is the gene product.  The distribution satisfies the chemical
master equation
\begin{eqnarray*}
    \frac{dp_0(n)}{dt} &=&  g_0p_0(n-1)-\left(g_0+nk\right)p_0(n)
\\
      &+& (n+1)kp_0(n+1)- h(n) p_0(n) + fp_1(n)
\\
    \frac{dp_1(n)}{dt} &=& g_1p_0(n-1)-\left(g_1+nk\right)p_0(n)
\\
      &+& (n+1)kp_0(n+1)+ h(n) p_0(n) - fp_1(n).
\end{eqnarray*}

We shall consider $h(n)=hn(n-1)/2$.
If the gene switching between $0$ and $1$ is slow, i.e., it is nonadiabatic,
then we have the rapid pre-steady states  (conditional probability) in the
state $i$ ($=0,1$) following their respective Poisson distribution
\begin{equation}
    p(n|i) = \frac{1}{n!}\left(\frac{g_i}{k}\right)^n e^{-g_i/k}.
\end{equation}
Then the mean transition rate from state $0$ to $1$, and $1$ to $0$, are
\begin{equation}
    k_{0\rightarrow 1} =  \sum_{n=0}^{\infty}h(n)
            p(n|0) = \frac{hg_0^2}{2k^2}, \ \ \
    k_{1\rightarrow 0} = \sum_{n=0}^{\infty} f p(n|1) = f.
\label{abs_nonadia}
\end{equation}

\subsection*{Perturbation method for eigenvalue problem} 

We can also
solve the eigenvalue problem using the method of perturbation
theory.  Note that the unperturbed system, i.e., when $h=f=0$, has
degeneracy.  For the perturbed system, we still have an eigenvalue
0, corresponding to the stationary distribution $[p(n|0),p(n|1)]^T$.
The other eigenvector on the left is $[1,-1]$ and on the right is
$[p(n|0),-p(n|1)]^T$, with the corresponding nonzero eigenvalue:
\begin{equation}
    \lambda_1 = -f -\frac{hg_0^2}{2k^2},
\label{lambda_1}
\end{equation}
which is exactly the $k_{0\rightarrow 1}+k_{1\rightarrow 0}$ in Eq.
\ref{abs_nonadia}. Expressing the $\lambda_1$ in terms of the
nondimensionalized parameters in [\cite{wolynes_pnas_05,shi_qian_11}], we have
\[
    -\frac{\lambda_1}{k} =\frac{h}{2k}\left(n_N^{\dag}\right)^2+
        \frac{hg_1^2}{2k^3}\left(\frac{g_0}{g_1}\right)^2
\]
\begin{equation}
    = \kappa \left[\left(\frac{k}{g_1}n_N^{\dag}\right)^2
        +\left(\frac{g_0}{g_1}\right)^2\right]
    = 0.29\kappa.
\end{equation}
The stochastic separatrix is well defined in this case by the
domains of positive and negative values of the eigenvector associated
with eigenvalue $\lambda_1$ [\cite{qian_jrsi_09}].

\subsection*{Nonadiabatic reduction to 1-d with $M$ 
copies of DNA}

The 1-d model has the transition rates:
\[
    m \rightarrow m+1:  \  \
    \frac{h_o(g_0(M-m)+g_1m)}{Vk}(M-m), 
\]
\begin{equation}
    m+1 \rightarrow m:  \ \
    f(m+1).
\end{equation}
If we let $x=m/V$ then we have
\begin{equation}
    (g_0-g_1)x^2+\left(g_1x_t-2g_0x_t-\frac{kf}{h_o}\right)x
        +g_0x_t^2 = 0.
\end{equation}
This should be compared with the quadratic equation
for the model in \cite{qian_bj_10}:
\begin{equation}
    (k_1+k_{-1})x^2-(k_1x_t-k_2-k_{-2})x-k_{-2}x_t = 0.
\end{equation}

\end{document}